\documentclass[12pt,epsf,a4paper]{article}
\usepackage[english]{babel}
\usepackage{latexsym,epsfig}
\usepackage{amsmath}

0

\begin{document}
\baselineskip 6,5mm

\def\II{\relax{\rm 1\kern-.35em1}}
\def\IP{\relax{\rm I\kern-.18em P}}
\renewcommand{\theequation}{\thesection.\arabic{equation}}
\csname @addtoreset\endcsname{equation}{section}

\begin{flushright}
hep-th/0004040 \\
IC/2000/32
\end{flushright}

\begin{center}

{}~\vfill

{ {\LARGE {Wilson Loops in the Higgs Phase of Large $N$ \\
Field Theories on the Conifold}}}

\vspace{20 mm}

{\bf {\large {Elena C\'aceres}}} {\large and} {\bf{\large{Rafael Hern\'{a}ndez}}}$^{\dag}$

\vspace{8 mm}

{\em The Abdus Salam International Center for Theoretical Physics \\
Strada Costiera, 11. $34014$ Trieste, Italy}
\vspace{16 mm}

\end{center}


\begin{center}
{\bf Abstract}
\end{center}
  
\vspace{2 mm}
  
We study the quark-antiquark interaction in the large $N$ limit of
 the 
superconformal field theory on D-3branes at a Calabi-Yau conical singularity. 
We compute the Wilson loop in the $AdS_5 \times T^{11}$ supergravity background for the 
$SU(2N) \times SU(2N)$ theory.
We also calculate the Wilson loop for the Higgs phase where 
the gauge group is broken to $SU(N) \times SU(N)\times SU_D(N)$.
This corresponds to a two center configuration 
with some of the branes at the singularity and the rest of them at a smooth point. 
The calculation exhibits the expected Coulomb dependence for the interaction. 
The angular distribution of the BPS states is different 
than the one for a spherical horizon.

\vspace{26 mm}

{\footnotesize \dag}\hspace{1 mm}{\footnotesize{\ttfamily e-mail address: caceres@ictp.trieste.it, 
rafa@ictp.trieste.it}}

\newpage


\section{Introduction}

The long suspected correspondence \cite{Gub,Pol} between string theory and the  
limit in which the number of colors of a gauge field theory is taken to 
infinity was first explicitly constructed 
by Maldacena \cite{ads/cft}, between a four dimensional $N=4$ supersymmetric gauge 
theory, and type IIB string theory compactified on $AdS_5 \times S^5$. According to the 
more precise statements of the conjecture elaborated in \cite{GKP} and \cite{Witten1}, 
correlation functions of gauge theory operators can be explicitly calculated from 
type IIB string theory on $AdS_5 \times S^5$. From the arguments in \cite{Witten1}, 
type IIB string theory on $AdS_5 \times X_5$, with $X_5$ a five dimensional Einstein 
manifold bearing five-form flux, is supposed to be dual to a four dimensional conformal 
field theory. Generalizations of the original idea have been constructed considering the sphere 
of the maximally supersymmetric case, $S^5$, divided by the action of some finite group 
\cite{Kachru}-\cite{Ber}. The field theory thus obtained corresponds to the infrared limit 
of the worldvolume theory on D-3branes at an orbifold singularity (the case of 7-branes 
at an orientifold singularity has also been considered in \cite{Kaku}-\cite{Aha}). Another 
proposal of duality arises from the study of D-3branes at Calabi-Yau singularities 
\cite{Keha}-\cite{KW2}. 
The connection between compactification on Einstein manifolds and the metric of D-3branes 
placed at a Calabi-Yau singularity pointed out in \cite{Keha}, was enlightened by 
Klebanov and Witten, who constructed the field theory side 
for a smooth Einstein manifold with local geometry different from the sphere, $X_5 
= T^{11}$ \cite{KW1}; their proposal is that the type IIB string theory compactified on 
$AdS_5 \times T^{11}$ should be dual to a certain large $N$ superconformal 
field theory with $N=1$ supersymmetry in four dimensions, which turns out to be a non trivial 
infrared fixed point, with 
gauge group $SU(N) \times SU(N)$. The field theory is constructed locating a collection of 
$N$ coincident D-3branes at a conical singularity of a non compact Calabi-Yau manifold (see 
also \cite{Morrison} for an exhaustive study of non-spherical horizons, and the 
corresponding field theories). In a subsequent paper \cite{KW2}, these authors also 
considered the posibility that the 
branes are extracted from the conifold point. Assuming that the branes are all at the same 
point, the low energy effective field theory will be the $N=4$ supersymmetric 
$SU(N)$ gauge theory, which is 
again an infrared fixed point. 
  
In this paper, we will consider an intermediate situation 
in between these two theories, in order to clarify the symmetry breaking, corresponding to 
a two center configuration, where some of the branes are still located at the conifold point, 
while the rest have been moved to some smooth point. We will study some details of 
this configuration 
describing the Higgs phase of the field theory of D-3branes at the conifold. 
Applying the construction for the Wilson loops introduced 
by Maldacena in \cite{Maldacena} to an $AdS_5 \times T^{11}$ background, 
we will explicitly calculate the energy for a quark-antiquark 
pair, separated by a distance $L$. As we will see, the 
form of the calculation we present also reduces naturally to the possibility that all branes 
are located at the conifold point, originally considered in \cite{KW1}, where we can also 
reproduce the expected conformal behaviour for the energy of the quark-antiquark pair. The 
expression for the energy, obtained minimizing the area of a string worldsheet in the 
$AdS_5 \times T^{11}$ background, allows also the possibility to understand configurations 
of vanishing energy.
  
In section 2, after a small review on the conifold and its properties, we will, following 
\cite{KW1}, study the $SU(N) \times SU(N)$ gauge theory broken 
to the Higgs phase $SU(N) \times SU(N) \times SU_D(N)$. The Higgs phase 
is explored in section 3 through the calculation of the Wilson loop in $AdS_5 
\times T^{11}$, for a configuration of $N$ branes at the conifold point, and $N$ branes located 
at a smooth point. The general solution we construct allows the possibility to consider 
the unbroken phase, with all branes located at the conifold. In section 4 we present 
some concluding remarks and possible implications 
of the present work, as well as some immediate steps further.


\section{Large $N$ Field Theories and Conifolds}

In this  section we will review some basic properties of 
a six dimensional Calabi-Yau manifold $Y_6$ with a conical singularity and study 
the Higgs branch for a particular configuration of D-3branes  
on $M_4 \times Y_6$. According to \cite{KW1} and \cite{KW2} keeping some of the branes at the 
conifold point implies $N=1$ supersymmetry. The configuration we will consider breaks the gauge 
group to $SU(N) \times SU(N) \times SU_D(N)$ and has $N=1$ supersymmetry.

\subsection{Conifold Spaces}
\label{conifold}
Let $Y_{6}$ be a six dimensional manifold smooth apart
from an isolated 
conical singularity, and of vanishing first Chern class.
$Y_6$ can be described by a quadric in 
${\bf C}^{4} \sim {\bf R}^{8}$,
\begin{equation}\label{e:conifold}
\mathcal{C} \equiv \sum_{A=1}^{4}\;\; (w^A)^2 =0.
\end{equation}
The singularity is located at $w^A=0$
for which ${\cal C}=0$ and $d{\cal C}=0$. Let $\mathcal{B}$ be the base of the cone.
$\mathcal{B}$ is given by the intersection of (\ref{e:conifold}) and a sphere of radius $\rho$ 
in ${\bf C}^4$,
\begin{equation}\label{e:sphere}
\sum_{A=1}^{4}\; |w^A|^2\;=\rho^2.
\end{equation}
  
In order to examine the topology of $\mathcal{B}$ we rewrite equations 
(\ref{e:conifold}) and (\ref{e:sphere}) in terms of the real 
and imaginary parts of $w^A$, $w^A= x^A +i y^A$, with $x^A,y^A \in {\bf R}$,
\begin{eqnarray}
x \cdot x & = & 1/2 \: \rho^2, \nonumber \\ 
y \cdot y & = & 1/2 \: \rho^2, \nonumber \\ 
x \cdot y & = & 0.
\end{eqnarray}
The first equation describes an $S^3$. The other two define a 
trivial  $S^2$ fiber over $S^3$. 
Thus, the base  $\mathcal{B}$ is topologically $S^2 \times S^3$ .

We want $Y_6$ to have vanishing first Chern class; however, 
the metric on a cone is Ricci flat if and only if the metric
on its base is Einstein; thus, $\mathcal{B}$ must be a 
5 dimensional Einstein
space, $\mathcal{R}_{\mu \nu} = 4 g_{\mu \nu}$ , with the topology of $S^2 
\times S^3$.
  
Let us now consider metrics on manifolds $T^{p,q}$ \footnote{The notation 
differs slightly from that in \cite{Candelas}, where these spaces are denoted by $N_{p,q}$.}
which are fiber bundles over 
$S^2 \times S^2$, with $(p,q)$ U(1) fiber, whose metric is \cite{Candelas}
\[
d\Sigma^2_{pq} = \lambda^2 ( d\psi + p\cos \theta_1 d\phi_1 + q \cos \theta_2 d\phi_2)^2 + 
\Lambda_1^{-1} (d\theta_1^2 + \sin^2 \theta_1 d\phi_1^2 )
\]
\begin{equation}\label{e:npqmetric}
+ \Lambda_2^{-1} (d\theta_2^2 + \sin^2 \theta^2 d\phi_2^2),
\end{equation}

\noindent where $0\le \theta_i < \pi$ and $0 \le \phi_i < 2\pi$
are spherical coordinates and $0 \le \psi <4 \pi$ is the coordinate 
on the U(1) fiber.
For some choices of $\Lambda_1, \Lambda_2,\lambda$ these metrics are 
Einstein. For two choices of $(p,q)$, namely $(1,1)$ and $(1,0)$, the fiber bundles 
are $S^2 \times S^3$. It was shown in \cite{Candelas} that only $T^{1,1}$ 
is compatible with a K\"ahler structure on the cone. In other words, 
these two Einstein metrics on the base give rise to two Ricci flat metrics
on the cone, but only taking $T^{1,1}$ as the base we get a cone which is  
a limit of a Calabi-Yau metric. In this case $Y_6$ will admit Killing spinors 
and thus, preserve supersymmetry. More precisely, $Y_6$ is a  
manifold of $SU(3)$ holonomy, therefore it preserves a 
quarter  of the supersymmetries of the 
original theory. This implies that 
the four dimensional theory will have 8 supercharges and  the theory will be  
$N=1$ supersymmetric. 

Redefining the radial coordinate as $r \equiv \sqrt{3/2} \rho^{2/3}$
we can write the $Y_6$ metric in the conical form,
\begin{equation}\label{e:conemetric}
ds^2_6=dr^2 + r^2d\Sigma^2_{11},
\end{equation}
where the metric of the base is 
\[
d\Sigma^2_{11} = \frac{1}{9} ( d\psi + \cos \theta_1 d\phi_1 + 
\cos \theta_2 d\phi_2)^2 + \frac{1}{6}(d\theta_1^2 + \sin^2 \theta_1 d\phi_1^2 )
\]
\begin{equation}\label{e:11metric}
 + \frac{1}{6}(d\theta_2^2 + \sin^2 \theta^2 d\phi_2^2).
\end{equation} 
The K\"ahler potential, $K$, for the conifold  metric (\ref{e:conemetric}) should 
be invariant under the $SU(2) \times SU(2)$ that acts 
on the base. In \cite{Candelas} the authors found  
 $K=(\sum_{A=1}^4 |w^A|^2)^{2/3}$.
  
For $T^{11}$ the $U(1)$ fiber is symmetrically embedded in the two 
$SU(2)$'s. The $U(1)$ generator is the sum  
$\frac{1}{2}\sigma_3 +\frac{1}{2}\tau_3$ of the $SU(2)$'s generators and  
$T^{11}$ is the coset space
\begin{equation} 
T^{11}=\frac{SU(2)\times SU(2)}{U(1)}=\frac{S^3 \times S^3}{U(1)}.
\label{t11}
\end{equation} 
\noindent This space can also be thought of as a smooth deformation 
of a blown-up $S^5/Z_2$ orbifold. The isometry group of the orbifold is $SO(4) \times SO(2)$, 
which is the same as for  $T^{11}$, and an 
appropiate blow up of the orbifold singularities leads to a manifold with the 
same topology as $T^{11}$.

\subsection{The Field Theory on D-3branes at the Conifold}

In \cite{KW1} the authors studied the theory obtained when
N D3-branes are placed on a conifold singularity.
They considered a decomposition of ten dimensional space given by $M_4 \times Y_6$, 
where $M_4$ is four dimensional Minkowski space, and $Y_6$ is 
a Calabi-Yau manifold with a conical singularity, of the form considered in section 2.1. Let 
the singularity be at a point $P$ on the conifold.
If we place $N$ parallel D3 branes on $M_4 \times P$, the 
resulting ten dimensional metric will be
\begin{equation}\label{e:generalcy}
ds^2= H(r)^{-1/2}[-dt^2 +dx^2] + 
H(r)^{1/2}[dr^2 +r^2g_{ij}dx^idx^j].
\end{equation}
\noindent where the harmonic function is $H(r)=1+\frac{L^4}{r^4}$, with 
$L^4=4\pi g_s N (\alpha ')^2$.
   
In the near horizon limit this metric becomes $AdS _5 \times T^{11}$,
where $T^{11}$ has been defined in (\ref{t11}), with the metric (\ref{e:11metric}). 
The infrared limit of the gauge theory on the D3 branes 
is $N=1$ supersymmetric with gauge group $SU(N)\times SU(N)$ 
and chiral superfields $A_i, B_j \ \ i,j = 1,2 $ . $A_i$ transforms as $({\bf N},\bar {\bf N})$ and 
$B_i$ as $(\bar {\bf N}, {\bf N})$ under  $SU(N)\times SU(N)$. The theory  has a non-renormalizable 
superpotential,
\begin{equation}\label{e:wsuperpotential}
W= \frac{\lambda}{2} \epsilon^{ij} \epsilon^{kl} A_iB_kA_jB_l.
\end{equation}
 
The moduli space of vacua of the theory is the conifold ${\cal C}$ defined in (\ref{e:conifold}).
The eigenvalues of $A_i, B_j$ are related to the the positions of the branes.
 
\vspace{4mm}
\noindent{\bf The Higgs Phase}
\vspace{2mm}

Symmetry breaking arises when the branes are moved away from the 
conifold singularity, away from each other or when the 
singularity is resolved or deformed \cite{KW1}. 
In the present paper we are interested in symmetry breaking 
obtained by moving the branes away from the conifold 
singularity. 
The case when all the branes are moved to a smooth point 
has been thoroughly studied. If the gauge symmetry is broken to 
the diagonal group
 \cite{KW2}, $SU(N) \times SU(N) \rightarrow SU_D(N)$,
the fields will transform in the adjoint representation of the group. 
One of them can be Higgsed away, leaving three adjoints 
with the superpotential 
\begin{equation}\label{e:n4superpotential}
W=\hbox{Tr}(\Phi_1[\Phi_2,\Phi_3]).
\end{equation}
 
This theory flows in the infrared 
to $N=4$ supersymmetric Yang Mills, as
expected since now the D3-branes are located on a 
smooth point of the manifold.  
  
But we want to investigate  a different 
pattern of symmetry 
breaking. We start with 2N branes at the conifold singularity,
{\it i.e.}, with a  $SU(2N)\times SU(2N)$ theory with two copies 
of matter in the $({\bf 2N},\overline{{\bf 2N}})$ and $(\overline{{\bf 2N}}, {\bf 2N})$ and 
a superpotential as described in equation (\ref{e:wsuperpotential}). We will consider 
the theory obtained when we move half of the branes, $N$ from each group, to a 
smooth point $\vec{r}=\vec{r}_0$ away 
from the singularity and leave the other half at 
the singularity $r=0$. The gauge group is then broken to $SU(N)\times SU(N)\times SU_D(N)$. 
The states charged under the $SU_D(N)$ and
either of the $SU(N)$'s correspond to  
strings stretching from the branes in the singularity to the branes 
at $r=r_0$, away from the conifold point. These are the W bosons 
responsible for the symmetry breaking. The fields left will transform as $({\bf N},
\bar{{\bf N}}, {\bf 1})$, $(\bar{{\bf N}},{\bf N},{\bf 1})$ and $({\bf 1},{\bf 1},{\bf N^2-1})$ 
under the Higgsed gauge group. Note that this pattern of symmetry breaking, as opposed to the one 
considered in \cite{KW2}, does not alter the amount of supersymmetry since the harmonic function 
is still singular at the conifold point $r=0$.
  
The process of moving branes out of the conifold 
could be done in steps, {\it i.e.}, successively 
moving groups of $M_i$ branes. In this case 
the gauge group 
at the smooth point will not 
be $SU_D(2N)$ but a collection of independant 
$SU_D(M_i)$'s. This can be understood as
a consequence of the different pattern of symmetry breaking, 
since  the intermediate steps imply 
the Higgsing of more states.


\section{Wilson Loops in Non Spherical Horizons}

The conjectured equivalence proposed by Maldacena between the large $N$ limit of 
four dimensional Yang-Mills with $N=4$ supersymmetry, and ten dimensional supergravity, allows 
the possibility to predict a number of interesting phenomena of field theories in the large $N$ 
limit. In \cite{Rey} and \cite{Maldacena}, a method to calculate the expectation value of 
the Wilson loop operator in the large $N$ limit of field theories was proposed 
(see \cite{Sonnenschein} for a recent review, and a collection of references on the subsequent 
cases studied in the literature). In this work, we will apply the technique introduced 
in \cite{Maldacena} to calculate the area of a fundamental string worldsheet in the 
$AdS_5 \times T^{11}$ background, and understand the energy, of a 
quark-antiquark pair, in the Higgs phase of the theory of branes at the conifold described 
in the previous section. The calculation we present also reproduces naturally the unbroken 
Higgs phase, where all branes are located at the conifold singularity.

\subsection{Multicenter Configurations}

The original configuration first studied by Maldacena in \cite{ads/cft} was 
a collection of $N$ parallel D-$3$branes, located at some point in their 
transverse space. The possibility to consider separated groups of D-3branes, 
that for large $N$ would also reproduce some Anti-de Sitter geometry, was also first 
mentioned in \cite{ads/cft}. Multicenter D-3brane solutions were originally 
considered in \cite{Duff}, and studied in the background of a Calabi-Yau threefold 
with a conical singularity in \cite{KW2}. The multicenter configuration corresponds to 
the harmonic function
\begin{equation}
H(\vec{r}) = 1 + \sum_a \frac { 4 \pi g_s N_a {\alpha '}^2}{|\vec{r}-\vec{\hat{r}}_a|^4},
\label{b1}
\end{equation}
where $\vec{r}$ denotes the coordinates transverse to the worldvolume of the D-3branes, 
$\vec{r}=\{y^1, \ldots, y^6 \}$, on the Calabi-Yau manifold, and $\vec{\hat{r}}_a$ are 
the positions of the D-3branes. We will measure distances, with respect to the conifold 
point, in terms of $r=\sqrt{y^{i} y^{i}}$. As in \cite{ads/cft}, the limit where 
$\alpha' \rightarrow 0$, keeping $r/\alpha'$ finite and all $N_a$ large, will 
correspond to the Higgs phase of an $SU(N)$ gauge theory, with $N= \sum_a N_a$, 
where the gauge group has been broken to $\prod_a SU(N_a)$. However, at large values 
of $r$ compared to the distances between the groups of D-3branes, which represent the Higgs 
vacuum expectation values, the multicenter configuration reproduces the Anti-de Sitter 
solution corresponding to $SU(N)$. As long as we keep some branes at the conifold point, as 
in section 2, the theory will be $N=1$ supersymmetric.
  
Following the construction in section 2, we will study the Wilson loop for a Higgs phase of 
the $N=1$ supersymmetric field theory of D-3branes at the conifold. The multicenter configuration 
will correspond to a set of N D-3branes located at some smooth point $\vec{r} = \vec{\hat{r}}_0$, and 
a collection of $N$ of them at the conifold point, so that the final gauge group is 
$SU(N) \times SU(N) \times SU_D(N)$, as in section 2. In the near horizon limit 
($\alpha ' \rightarrow 0$ but $r/\alpha '$ finite), the harmonic function becomes
\begin{equation}
H(\vec{y}) = \frac {4 \pi g_s N {\alpha '}^2}{r^4} + \frac {4 \pi g_s N {\alpha '}^2}
{|\vec{r} - \vec{\hat{r}}_0|^4}.
\label{b2}
\end{equation}
  
For the harmonic function (\ref{b2}), there is a clear limit, corresponding to 
$r \gg \hat{r}_0$, where the expression for $H$ simplifies to that of a single collection 
of $2N$ D-3branes located at the conifold point, $H= \frac {2 \; . \;4 \pi g_s N {\alpha'}^2}{r^4}$. 
We will also address this limit for our calculation at the end of this section. 
From a field theory point 
of view, it implies that the mass scale of the symmetry breaking of $SU(2N) \times SU(2N)$ 
to $SU(N) \times SU(N) \times SU_D(N)$, proportional to the distance in between the 
two sets of branes, $r_0$, is much smaller than the energy scale of the quark-antiquark 
potential, so that the interaction will correspond to the unbroken $SU(N) \times SU(N)$ phase.
  
We are interested in the harmonic function (\ref{b2}) describing the configuration with two 
centers. However, from a calculational point of view, this function is extremely 
involved. In fact, a similar 
calculation for a Higgs phase of branes at two different positions, but for the 
$AdS_5 \times S^5$ background, was considered in 
\cite{Minahan}. The authors of \cite{Minahan} used a symmetric configuration in order 
to simplify the form of the harmonic function. Our work requires a similar choice of 
symmetric configuration\footnote{Another possibility to describe the harmonic function of a multicenter 
configuration in an $AdS_5 \times T^{11}$ background is to use an expansion for $H$ in terms of 
spherical harmonics, as in \cite{KW2}. However, again from a calculational point of view, the 
harmonic function leads to extremely complicated expansions for the solutions, which can only 
be calculated as a power series.}. The main difficulty arises due to the fact 
that $H$ contains two terms 
for a two center configuration, and that it is $H^{-1}$ that enters the 
Nambu-Goto action in the calculation of the Wilson loop. Hence, the most convenient possibility 
is to restrict the calculation to the region where 
\begin{equation}
|\vec{r}|=|\vec{r}-\vec{\hat{r}}_0|.
\label{symcon}
\end{equation}
This constraint corresponds to a symmetric configuration where the minimum of the Wilson loop 
is located at the vertex of the distinct angle of an isosceles triangle. The 
two sets of branes (those located at the conifold, and those at the smooth point) are then at the two 
other vertices, so that the two equal sides of the triangle will correspond to the distance 
to the loop. 
  
As we will see next, the Wilson loop can be calculated exactly for this symmetric configuration.

\subsection{Geodesics of the Symmetric Configuration}

In order to calculate Wilson loops according to the prescription in \cite{Maldacena}, 
the Nambu-Goto action of a fundamental string should be minimized in a relevant 
supergravity background. We should then consider the action for the string worldsheet,
\begin{equation}
S = \frac{1}{2 \pi \alpha '} \int d \sigma d \tau \sqrt{\hbox {det } G_{MN} 
\partial_{\alpha} X^M \partial_{\beta} X^N},
\label{a1}
\end{equation}
in the $AdS_5 \times T^{11}$ background, where the $G_{MN}$ metric is
\begin{equation}
d s_{10}^2 = H^{-1/2}(r) (-dt^2 + dx_i dx^{i}) + H^{1/2}(r) (dr^2 + r^2 d \Sigma_{11}^2).
\label{a2}
\end{equation}
Defining $U = r/\alpha'$, this metric becomes
\begin{equation}
d s_{10}^2 = \alpha' [ H^{-1/2}(U) (-dt^2 + dx_i dx^{i} ) + H^{1/2}(U) 
(dU^2 + U^2 d \Sigma_{11}^2)],
\label{a3}
\end{equation}
so that now the harmonic function describing the multicentered configuration is
\begin{equation}
H(U) = \frac {R^4}{U^4} + \frac {R^4}{|\vec{U}-\vec{r}_0|^4},
\label{a4}
\end{equation}
where we have introduced $R^4 \equiv 4 \pi g_s N$, and the rescaling $\vec{r}_0 \equiv 
\vec{\hat{r}}_0 /\alpha'$ has also been employed, so that $\vec{r}_0$ is kept constant.
  
Now, following \cite{Maldacena}, we can introduce a relative angle between the quarks, 
by giving expectation values $\vec{\Phi}_1$ and $\vec{\Phi}_2$ to two $U(1)$ factors 
in the global gauge group. If the angles are defined as $\vec{\theta}_i = 
\vec{\Phi}_i / |\vec{\Phi}|$, the string worldsheet will be located, at the end of 
the Wilson loop corresponding to the position of the massive quark, at $U= \infty$ and 
a point $\vec{\theta}_1$ on the $T^{11}$ space, and at $U = \infty$ but a point 
$\vec{\theta}_2$ on $T^{11}$ for the other end of the loop, corresponding to the position 
of the other massive quark. However, as seen in section 2.1, the topology of $T^{11}$ is 
quite different from that of $S^5$. The $SO(6)$ isometry group of $S^5$ was responsible 
that the string joining $\vec{\theta}_1$ and $\vec{\theta}_2$ 
in \cite{Maldacena}, would simply lie on a great circle of the sphere. The $SO(4) 
\times SO(2)$ isometry group of $T^{11}$ also allows this possibility: the Wilson 
loop can be taken to extend along a great circle in $S^2$, or in $S^3$; this simply 
corresponds to setting all angles, but one, to constants, in the metric describing 
$T^{11}$. The loop constructed this way clearly reproduces that studied in \cite{Maldacena} 
for strings located at different points in $S^5$, up to a numerical coefficient related 
to the constraints defining the $T^{11}$ metric. However, $SO(4) \times SO(2)$ does not exclude 
other possibilities. The Wilson loop can be taken to lie on a path parametrized by two, or 
even three, different coordinates on $T^{11}$, and which cannot be described in terms of 
a single angle through a rotation. For definiteness, we will choose the path joining  
$\vec{\theta}_1$ and $\vec{\theta}_2$ to be parametrized by $\psi$ and $\phi_1$ (the 
two other cases with two angles, $(\psi,\phi_2)$ and $(\phi_1,\phi_2)$, or the 
slightly more involved of a trajectory requiring the set $(\psi,\phi_1,\phi_2)$, can be 
similarly treated). Then, setting $\phi_2 = \hbox {constant}$, and choosing the convenient 
values $\theta_1 = \theta_2 =0$, the action for a static configuration becomes
\begin{equation}
S = \frac {T}{2 \pi} \int dx \sqrt{H^{-1}(U) + (\partial_x U)^2 +  {U^2}/{9} 
(\partial_x \psi)^2 +  {U^2}/{9} (\partial_x \psi) (\partial_x \phi_1) +
{U^2}/{6} (\partial_x \phi_1)^2},
\label{a5}
\end{equation}
where we have identified the space worldsheet coordinate $\sigma$ with one of the spatial 
coordinates on the worldvolume of the D-3branes.
  
The Euler-Lagrange equations for this action are, in general, quite hard to solve for 
the two center configuration we are interested in, 
unless a symmetric constraint as the one described in 3.1 is chosen. We will work in 
such a configuration, so that our solutions will represent the geodesics satisfying 
condition (\ref{symcon}).
  
As the action does not depend on $x$ explicitly, the associated energy
\begin{equation}
\frac {U^4/2}{\sqrt{H^{-1}(U) + (\partial_x U)^2 + {U^2}/{9} 
(\partial_x \psi)^2 +  {U^2}/{9} (\partial_x \psi) (\partial_x \phi_1) +
{U^2}/{6} (\partial_x \phi_1)^2}}
\label{a6}
\end{equation}
will be a conserved quantity. Besides, the action is also cyclic in $\psi$ and $\phi_1$, 
so that the associated angular momenta, ${\partial L}/{\partial (\partial_x \psi)}$ and 
${\partial L}/{\partial (\partial_x \phi_1)}$ (where the lagrangian $L$ is the 
square root in (\ref{a5})), will also be constants. The value of these constants can be 
fixed in terms of the minimum of the Wilson loop. If $U_0$ is the distance along the 
slice transversal to the direction defined by the position of the conifold point, and the 
$N$ D-3branes at $\vec{r}_0$, outside the singularity\footnote{More geometrically, 
$U_0$ will correspond to the symmetric height of the triangle.}, then, in the symmetric configuration of 
section 3.1, the minimum will correspond to a radial coordinate $U_{\hbox{\tiny{min}}} = \sqrt{U_0^2 +
\left( \frac {r_0}{2} \right)^2}$, measured with respect to any of the two sets of D-3branes. 
Then, solving for $\partial_x \psi |_{U_{\hbox{\tiny{min}}}}$ and 
$\partial_x \phi_1 |_{U_{\hbox{\tiny{min}}}}$, the constants can be fixed to
\begin{eqnarray}
\frac {\partial L}{\partial (\partial_x \psi)}   & = & \sqrt{U_0^2 +
\left( \frac {r_0}{2} \right)^2} l_{\psi}, \nonumber \\
\frac {\partial L}{\partial (\partial_x \phi_1)} & = & \sqrt{U_0^2 +
\left( \frac {r_0}{2} \right)^2} l_{\phi_1},
\label{a7}
\end{eqnarray}
while 
\begin{equation}
(U_0^2 + \left( \frac {r_0}{2} \right)^2) R^2 \sqrt{1-f^2(l_{\psi}, l_{\phi_1})}
\label{a8}
\end{equation}
for the energy associated to $x$, where we have defined
\begin{equation}
f^2(l_{\psi}, l_{\phi_1}) = \frac {18}{5} (3 l_{\psi}^2 - 2 l_{\psi} l_{\phi_1} + 
2 l_{\phi_1}^2).
\label{a9}
\end{equation}
  
With the value of the angular momenta fixed through (\ref{a7}) and (\ref{a8}), 
and (\ref{a6}) for the energy, and using the fact that the minimum is located at 
$\sqrt{U_0^2 + \left( \frac {r_0}{2}
\right)^2}$, the solution of these equations for $x$, $\psi$ and $\phi_1$, in terms of $U$ is
\begin{eqnarray}
x      & = & \frac {R^2 \sqrt{1-f^2(l_{\psi},l_{\phi_1})}}{\sqrt{U_0^2 + 
\left( \frac {r_0}{2} \right)^2}} 
\int_1^{U/\sqrt{U_0^2 + \left( \frac {r_0}{2} \right)^2}} \frac {dy}{y^2 
\sqrt{(y^2-1)(y^2+1-f^2(l_{\psi},l_{\phi_1}))}}, \nonumber \\
\psi   & = & - \frac  {9(3 l_{\psi}-l_{\phi_1})}{10} \int_1^{U/\sqrt{U_0^2 + 
\left( \frac {r_0}{2} \right)^2}} \frac {dy}{y^2 
\sqrt{(y^2-1)(y^2+1-f^2(l_{\psi},l_{\phi_1}))}}  \nonumber \\
\phi_1 & = &   \frac {9 (l_{\psi} - 2 l_{\phi_1})}{5} \int_1^{U/\sqrt{U_0^2 + 
\left( \frac {r_0}{2} \right)^2}} \frac {dy}{y^2 
\sqrt{(y^2-1)(y^2+1-f^2(l_{\psi},l_{\phi_1}))}}.
\label{a10}
\end{eqnarray}
The value of the parameter $U_0$ can be determined through the condition
\begin{equation}
\frac {L}{2} = x(U = \infty) =\frac {R^2 \sqrt{1-f^2(l_{\psi},l_{\phi_1})}}{\sqrt{U_0^2 + 
\left( \frac {r_0}{2} \right)^2}} I_1(l_{\psi},l_{\phi_1})
\label{a11}
\end{equation}
where, as in \cite{Maldacena}, $I_1(l_{\psi},l_{\phi_1})$ represents an expression 
that can be calculated in terms of elliptic integrals,
\begin{equation}
I_1(l_{\psi},l_{\phi_1}) = \frac {1}{(1-f^2) \sqrt{2-f^2}} \biggl[ (2-f^2) E \left( \frac 
{\pi}{2}, \sqrt{\frac {1-f^2}{2-f^2}} \right) - F \left( \frac 
{\pi}{2}, \sqrt{\frac {1-f^2}{2-f^2}} \right) \biggr].
\label{a12}
\end{equation}

Similarly, $l_{\psi}$ and $l_{\phi_1}$ can be obtained from the conditions 
\begin{eqnarray}
\frac {\Delta \psi}{2} & = & \psi (U = \infty) = - \frac {9(3 l_{\psi}-l_{\phi_1})}{10} 
I_2(l_{\psi},l_{\phi_1}), \nonumber \\
\frac {\Delta \phi_1}{2} & = & \phi_1 (U = \infty) = \frac {9 (l_{\psi} - 2 l_{\phi_1})}{5} 
I_2(l_{\psi},l_{\phi_1}),
\label{a13}
\end{eqnarray}
where $I_2(l_{\psi},l_{\phi_1})$ is the integral
\begin{equation}
I_2(l_{\psi},l_{\phi_1}) = \frac {1}{\sqrt{2-f^2}} F \left( \frac {\pi}{2}, 
\sqrt{\frac {1-f^2}{2-f^2}} \right),
\label{a14}
\end{equation}
again as in \cite{Maldacena}, but now with $f$ defined in (\ref{a9}). 
  
Equation (\ref{a11}) exhibits 
the dependence of the radial position of the minimum, $U_{\hbox {\tiny{min}}}$, in the quark 
distance, but also of this $L$ in the scale $r_0$ introduced through Higgs mechanism when 
splitting the set of branes in the bulk. As the distance $r_0$ to the smooth point 
decreases, $r_0 \ll U_0$, from (\ref{a11}) we recover an inverse 
law for the relation between $L$ and $U_0$, as in the $AdS_5 \times S^5$ Wilson loop 
constructed in \cite{Maldacena}. The only difference with that case, as a consequence of the 
different background, is the $f^2$ function (\ref{a9}). Note also that the large separation 
limit, $r_0 \gg U_0$, leads to small values of $L$, as in the Higgs phase studied in \cite{Minahan}. 
  
The energy can be calculated if the solutions (\ref{a10}) are introduced in the expression 
(\ref{a5}) for the action. The infinite result, arising from the mass of the W boson 
corresponding to a string stretching to $U= \infty$, can be regularized 
introducing some cut-off, so that the energy is simply integrated up to some $U_{\hbox 
{\tiny{max}}}$ \cite{Maldacena}. When the regularized mass $U_{\hbox {\tiny{max}}}/2 \pi$ of the 
W boson is substracted, the energy
\begin{equation}
E = \frac {\sqrt{U_0^2 + \left( \frac {r_0}{2} \right)^2}}{2 \pi} \biggl[ \int_1^{\infty} 
dy \left( \frac {y^2}{\sqrt{(y^2-1)(y^2+1-f^2)}} -1 \right) -1 \biggr]
\label{a15}
\end{equation}
leads, once (\ref{a11}) is taken into account to introduce the distance $L$ separating the 
two quarks, to the result
\begin{equation}
E = - \frac {1}{\pi} \frac {(4 g_{YM}^2 N)^{1/2}} {L} (1-f^2)^{3/2} I_1^2(l_{\psi},l_{\phi_1}).
\label{a16}
\end{equation}
Thus, we see that for any value of the distance $L$, as compared to the scale 
$r_0$ measuring the W mass, we obtain a coulombic behaviour for the quark potential. The 
Coulomb dependence is the same as that obtained in \cite{Rey,Maldacena} for the unbroken 
Higgs phase of $N=4$ theories in the large $N$ limit. This dependence was also obtained for 
small values of $L$ in \cite{Minahan} for the Higgs phase of $N=4$ supersymmetric 
Yang-Mills, where the breaking to $SU(N/2) \times SU(N/2)$ was considered. Note that 
the potential (\ref{a16}) is 
coulombic even in the large $L$ region, where the energy scale is much smaller than the 
mass of the $W$ particles, which therefore become irrelevant; in this region, 
we should then still expect a coulombic dependence\footnote{Large values of $L$ modified 
the coulombic 
dependence of the energy in \cite{Minahan}; however, the authors argued the existence of other 
geodesics, beyond their symmetric choice, where the $L^{-1}$ behaviour should be restored.}.
  
We thus capture the expected Coulomb behaviour of the 
interaction energy in the limit of 
very large and very small quark-antiquark distance $L$. However, 
deviation from a Coulomb 
dependence is expected at intermediate distances. The energy scale is 
then comparable 
to the W boson mass which becomes relevant and corrects the 
interaction of the quarks. 
This behaviour cannot be obtained from geodesics obeying the 
symmetric constraint 
because once (\ref{symcon}) is imposed in the 
harmonic function the metric becomes exactly AdS. In order to study 
deviations 
from the conformally invariant behaviour indirectly imposed by the 
symmetric choice 
(\ref{symcon}), we should consider more general geodesics. The 
authors of \cite{Minahan}, when 
studying the Higgs phase of a two center configuration in $N=4$ 
supersymetric 
Yang-Mills, exhibited the validity of their symmetric choice of 
geodesics by studying 
small perturbations of their solution. Their approach turns however 
estremely involved 
in our case since we are trying to 
capture the features associated 
with a 
non spherical manifold, the $T^{11}$. However, most of their 
arguments should still be valid in our case.
   
In order to exhibit deviations from the Coulomb behaviour arising 
from non conformality at 
intermediate energies (or distances), we will calculate the 
expression for $x=x(U)$ in a 
configuration different from the symmetric one (the explicit form of 
this expression, as well 
as those for $\psi(U)$ and $\phi_1(U)$, can be obtained along the 
same lines as 
(\ref{a10})). As the calculation is rather involved, 
and requires a series expansion, we will chose to concentrate, for 
simplicity, in geodesics 
satisfying the radial requirement that both $\vec{U}$ and $\vec{r}_0$ 
are along the same 
direction, so that the harmonic function can be written as $H(U) = 
\frac {R^4 (U^4 
+ (U-r_0)^4)}{U^4(U-r_0)^4}$. Solving (\ref{a6}) and (\ref{a7}) for 
$\partial_x U$, we find that now the value of $U_{\hbox{\tiny{min}}}$ 
should be 
determined through the series
\begin{equation}
\frac {L}{2} = \frac {R^2}{U_{\hbox{\tiny{min}}}} \biggl[ F_0(l_{\psi},l_
{\phi_1}) + 
\left( \frac {r_0}{U_{\hbox{\tiny{min}}}} \right) F_1(l_{\psi},l_
{\phi_1}) + 
{\cal O} \left( \left( \frac {r_0}{U_{\hbox{\tiny{min}}}} \right) ^2 
\right) \biggr],
\label{nueva}
\end{equation}
where $F_0(l_{\psi},l_{\phi_1})$ and $F_1(l_{\psi},l_{\phi_1})$ are integrals that can 
be calculated in terms 
of elliptic functions. When $r_0 \ll U_{\hbox{\tiny{min}}}$, we can 
drop terms of order $\frac{r_0}{U_{\hbox{\tiny{min}}}}$ to recover a Coulomb 
dependence for the interaction energy of the 
quark-antiquark pair\footnote{Note that we simply present as argument for the deviation 
from conformality the relation between $L$ and $U_{\hbox{\tiny{min}}}$. The explicit form 
of $E=E(L)$ should be obtained graphically, but expression (\ref{nueva}) is enough 
to apreciate the origin of the non coulombic dependence.}. Besides, from an expansion in 
$\frac {U_{\hbox{\tiny{min}}}} {r_0}$ it can be shown that the Coulomb law dependence 
is also obtained in the opposite limit, where 
$r_0 \gg U_{\hbox{\tiny{min}}}$, as expected. However, higher order terms will correct the 
conformal behaviour, as 
in \cite{Minahan}. We thus see that the exact 
conformal behaviour exhibited by the symmetric configuration is corrected 
at intermediate distances by other geodesics, as this calculation along the radial direction 
shows.
    
Finallly, to conclude the analysis of the solution for the Higgs phase of branes at the 
conifold in the a symmetric configuration, we note that the value of the 
energy does not depend on the distance $r_0$ explicitly, so that 
expression (\ref{a16}) also represents the dependence of the energy in the un-Higgsed theory, when 
$r_0=0$ and all D-3branes are at the conifold point. This is also the result obtained in the 
$U \gg r_0$ limit.

\subsection{Analysis of the Trajectories}
  
The regularized value for the total energy of the configuration, (\ref{a16}), is formally the 
same as that found in \cite{Maldacena} for the case of two quarks at a non constant angle. The 
missing factor of $2$ is a consequence of the two center distribution of branes, 
constrained to the symmetric condition of section 3.2 \footnote{We have kept this factor 
explicitly in the definition of the energy in (\ref{a6}). Note also the factor $4$ in the square 
root in (\ref{a16}), instead of the usual $2$, as a consequence of the collection of $2N$ 
D-3branes that we are considering.}. However, the requirement of more than 
one angle to parametrize the Wilson loop on $T^{11}$ is responsible that $f^2$, entering the 
energy, becomes now a function of the angular momenta, as defined in (\ref{a9}). The vanishing energy 
of a BPS configuration is shown in \cite{Maldacena} to correspond to the limit 
$\Delta \theta \rightarrow \pi$, so that the orientation of the two strings is reversed. 
Now, the $f^2=1$ condition on the angular momenta, associated to the zero values of the 
energy, becomes the curve
\begin{equation}
\frac {18}{5} (3 l_{\psi}^2 -2 l_{\psi} l_{\phi_1} + 2 l_{\phi_1}^2) =1.
\label{a17}
\end{equation}
If we imposse $f^2=1$ in (\ref{a13}), condition (\ref{a17}) can be written in terms of the 
angles $\psi$ and $\phi_1$, parametrizing the loop, as
\begin{equation}
\frac {2}{3} \frac {(\Delta \phi_1)^2}{\pi^2} - \frac {8}{9} \frac {\Delta \phi_1 
\Delta \psi}{\pi^2} + \frac {16}{9} \frac {(\Delta \psi)^2}{\pi^2} =1,
\label{a18}
\end{equation}
which is the equation of a rotated ellipse in the $(\psi,\phi_1)$ plane. This curve generalizes 
the condition found in \cite{Maldacena}, where trajectories associated to a variation 
$\Delta \theta = \pi$ of the angle measuring the relative orientation of the two strings 
representing the quarks were shown to correspond to a BPS configuration. Now, all trajectories 
such that the path where the loop is contained corresponds to a set of values of 
$\psi$ and $\phi_1$ in the ellipse (\ref{a18}), are trajectories with vanishing energy. 
Besides, all trajectories whose variation along the $\psi$ and $\phi_1$ directions 
correspond to points inside the ellipse (points such that $f^2 \leq 1$), including 
the origin $\psi=\phi_1=0$, where no relative angle has been introduced, will 
lead to real values for the energy.  
  
We thus see how the topology of the non spherical horizon $T^{11}$ does not require 
reversed orientation of the strings, corresponding to $\Delta \theta = 
\pi$ in the $AdS_5 \times S^5$ background \cite{Maldacena}.


\section{Conclusions}

Previous work in the Higgs \cite{Minahan} and Coulomb branches
\cite{larsen}
 of $N=4$ $SU(N)$
 Yang-Mills  dual to 
$AdS_5 \times S^5$ has proven very fruitful. In particular,
continuous distributions of branes \cite{larsen}-\cite{sfetsos3} provide
examples of renormalization 
group flow in an AdS/CFT setting \cite{Alvarez}-\cite{gubser3}.
In the present work we began 
a program to study similar examples of continuous brane distributions 
for theories with non spherical 
horizons \cite{nosotros}. As a first step we have calculated the energy for 
a quark-antiquark pair in an $SU(N)\times SU(N) \times SU_D(N)$
theory with N=1 supersymmetry using the richer topology of $T^{11}$ background. 
This  Higgs branch corresponds to
the flow of the original $SU(2N)\times SU(2N)$ field theory under 
symmetry breaking. 
We found the expected coulombic behaviour for the quark-antiquark
potential, because the harmonic function in the symmetric configuration that 
we are using still has a quartic dependence on the radial coordinate; thus, we see 
that the general results obtained in \cite{Gi} also hold for backgrounds with non spherical 
horizons. As a by product we also obtain the energy of a quark-antiquark pair 
for the un-Higgsed theory. The energy for the configuration with 
all branes at the 
smooth point cannot be obtained from our result, and requires a series 
expansion. Besides, the relative angle in between the two strings allows 
the possibility to study states with vanishing energy. For both Higgsed and un-Higgsed cases 
we find that unlike the $AdS_5\times S^5$ background, there exists a two dimensional 
curve of BPS states, obtained from different orientation of the quarks, whose existence 
seems to be a characteristic of non spherical horizons theories. However, 
the interpretation in the field theory side of these states, obtained from strings with  
relative orientation different from $0$ or $\pi$, deserves further study.


\vspace{8 mm}

{\bf Acknowledgements}

It is a pleasure to thank E. Gava, K. Narain and K. Ray for 
useful discussions, and J. Maldacena for comments. This research is 
partly supported by the EC contract no.
ERBFMRX-CT96-0090.

\newpage


\end{document}